\documentclass[10pt,letterpaper,twocolumn]{article} 

\usepackage{ol2}
\usepackage[draft,implicit=false]{hyperref}
\usepackage{amsmath}

\newcommand{\ion}[2]{\mbox{$^{#2}$#1$^+$}}
\newcommand{\Ca}[1]{\ion{Ca}{#1}}

\newcommand{\CaII}[1]{\ion{Ca}{#1}}

\begin{document}

\twocolumn[ 

\title{Injection locking of two frequency-doubled lasers with $3.2\:$GHz offset for driving Raman transitions with low photon scattering in $^{43}$Ca$^+$ }


\author{N. M. Linke,$^{*}$ C. J. Ballance, and D. M. Lucas}

\address{
$^1$Clarendon Laboratory, Department of Physics, University of Oxford, Parks Road, Oxford OX1 3PU, UK
\\
$^*$Corresponding author: linke@physics.ox.ac.uk
}

\begin{abstract}
We describe the injection locking of two infrared ($794\:$nm) laser diodes which are each part of a frequency-doubled laser system. An acousto-optic modulator (AOM) in the injection path gives an offset of $1.6\:$GHz between the lasers for driving Raman transitions between states in the hyperfine split (by $3.2\:$GHz) ground level of $^{43}$Ca$^+$. The offset can be disabled for use in \Ca{40}. We measure the relative linewidth of the frequency-doubled beams to be $42\:$mHz in an optical heterodyne measurement. The use of both injection locking and frequency doubling combines spectral purity with high optical power. Our scheme is applicable for providing Raman beams across other ion species and neutral atoms where coherent optical manipulation is required. 
\end{abstract}


 ] 

\noindent 
Ions held in rf traps are a standard system for quantum information processing experiments \cite{Wineland11}. One of the key challenges is to realize coherent operations with fidelities high enough to allow the implementation of quantum error correction protocols \cite{Rauss12}. The sub-states of the ground level of alkali earth ions are commonly used as the qubit states \cite{Home2006c,Itano98}. In \CaII{43} with its nuclear spin ($I=7/2$) a pair of Zeeman states from the $F=3$ and $F=4$ hyperfine manifolds can be chosen \cite{Lucas07,Benhelm2008b}. At specific magnetic field strengths pairs of magnetic-field insensitive clock-states are available that have recently been shown to have extremely long coherence times \cite{Allcockpaper} making \CaII{43} a promising species for quantum information. 

Coherent transitions between the qubit states may either be driven by rf/microwave radiation \cite{AllcockMW12} or by a pair of Raman laser beams near the $4S_{1/2}$ to $4P_{1/2}$ transition at $397\:$nm \cite{Itano98}. Their strong coupling to the motion makes Raman beams particularly attractive for sideband cooling and motional entangling gates. 

With Raman lasers the choice of detuning $\Delta$ from the $4S_{1/2}$ to $4P_{1/2}$ transition is a compromise between higher speed at small detuning (scaling $\sim\,1/\Delta$) and lower off-resonant photon scattering (which introduces qubit errors) at large detuning (scaling $\sim\,1/\Delta^2$). Typical detunings are in the range of $10-1000\:$GHz. For a desired gate speed, the fidelity of qubit operations depends on the available Raman beam power. Typically, powers around $1\:$mW have been used. For common experimental parameters and a desired two-qubit gate error probability of $10^{-4}$ in a gate time of $10\:\mu$s, beam powers of around $100\:$mW are required \cite{Kni08,Ozeri07}.

The two-photon Raman process only depends on the difference frequency between the two beams and its detuning from the qubit resonance which is why there are no stringent requirements for absolute frequency stability. Instead, relative interferometric stability between the beams is needed. The hyperfine ground state splitting in \CaII{43} is 3,225,608,286.4(3)\:Hz \cite{arbes94}. This can be spanned by using an acousto-optic modulator (AOM) in one of two beams derived from a single source but the low diffraction efficiency ($\sim10\%$ double-pass) and damage threshold ($\sim1\:$mW) make this approach unattractive. More efficient electro-optic modulators (EOM) are available but here the upper and lower sidebands remain colinear with the carrier and separating them spatially is not straightforward and entails a cost in optical power. Despite recent progress \cite{koda10}, tapered amplifiers (TAs) are not yet commercially available at $397\:$nm. Another alternative is injection locking, where only a low intensity seed beam needs to be frequency shifted. This has recently been reported with violet diode lasers \cite{Keitch13}. While this approach is currently limited to $\sim20\:$mW of output power, it shows promise as higher power ($120\:$mW) diodes have recently become available.

The approach we implement in this Letter makes use of the established techniques of injection locking \cite{spano85,Bouyer96} and the availability of TAs in the infrared \cite{Walpole92} to provide two coherent beams of $450\:$mW intensity at $794\:$nm. We use a double-pass $800\:$MHz AOM\footnote{Brimrose TEF-800-300-.794} in the injection path to provide a relative frequency offset of $1.6\:$GHz. The beams are then frequency-doubled to give $120\:$mW of violet light in each, with the required frequency difference of $3.2\:$GHz. Apart from the much increased power the scheme has an additional advantage over direct injection in the violet. The frequency-doubling cavity acts as a spectral filter which prevents near-resonant light from amplified spontaneous emission (ASE) in the laser diode or the TA from reaching the ion and causing unwanted photon scattering. We confirmed the spectral purity of the violet light in a measurement of the photon scattering rate of a single ion for different Raman detunings. Injection locking gives the flexibility of making the injection path switchable between the first and zeroth order of the AOM allowing the system to be used both with a $3.2\:$GHz offset for \CaII{43} and without for \CaII{40} (in the latter case the few MHz offset can be provided through subsequent switching AOMs at $397\:$nm). We show a high level of coherence between the two violet beams in an optical heterodyne measurement. An analogous setup would be applicable to different ion species or neutral atoms.

\begin{figure}[t]
\centerline{\includegraphics[height=6.6cm]{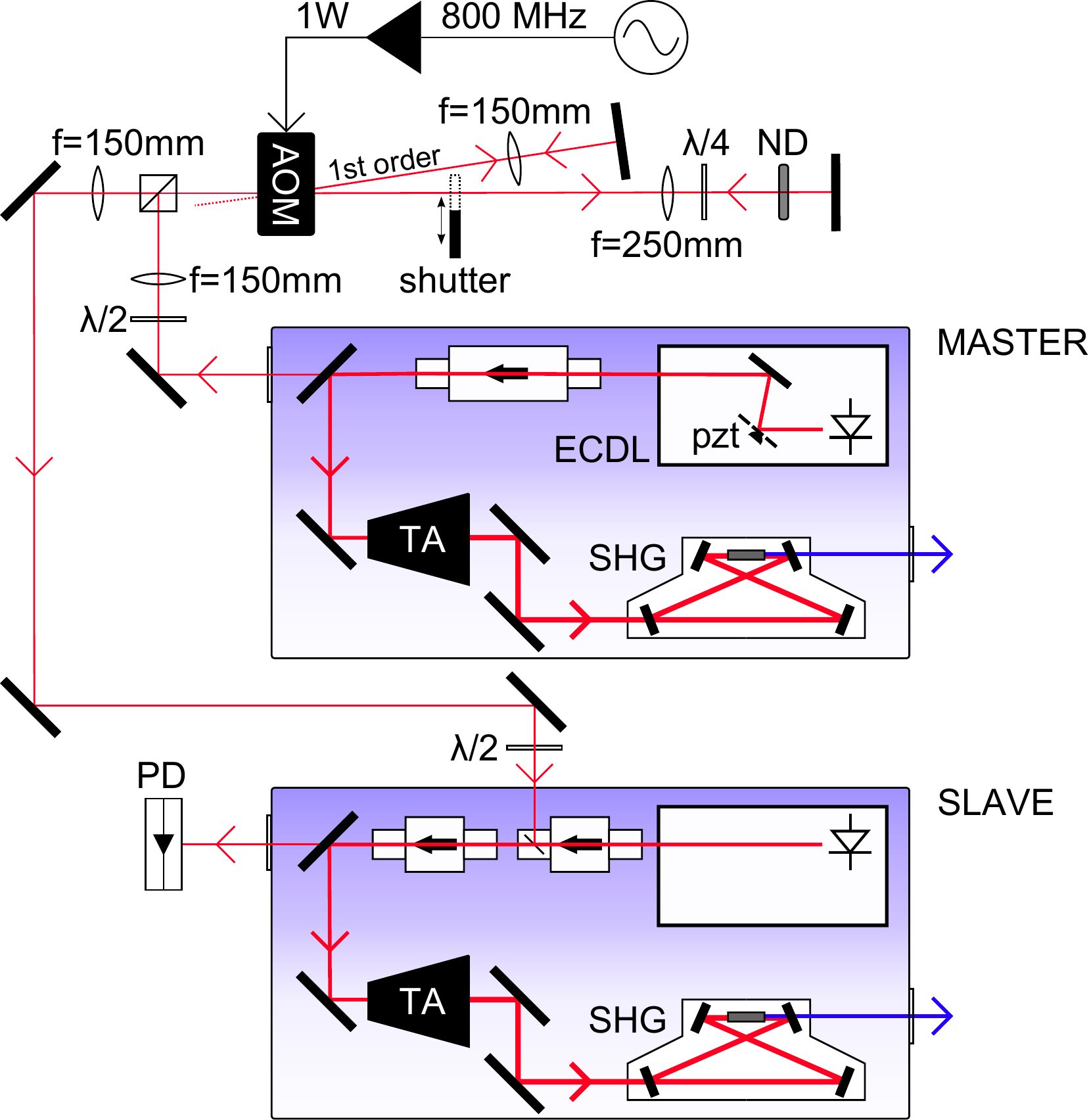}}
\caption{Injection beam setup showing the double-pass AOM arrangement which allows the offset to be switched between $1.6\:$GHz (first order injected, zeroth order blocked) and $0\:$GHz (AOM off, zeroth order unblocked). Both master and slave are amplified and frequency-doubled. See text for key.}
\label{injectionsetup}
\end{figure}

The laser systems are commercial extended-cavity diode lasers (ECDLs) with TA and second-harmonic generation (SHG) stages\footnote{Master: Toptica TA/DL-SHG 110 Pro; Slave: Toptica TA/DL-SHG 110 with AR coated diode}. Originally, both were grating-stabilized in Littrow configuration. The linewidth of the free running systems was estimated in a heterodyne experiment to be $300\:$kHz. The grating on the slave was subsequently removed (see fig. \ref{injectionsetup}). The master beam power after the $30\:$dB Faraday isolator is typically $35\:$mW. A weak beam ($5\:$mW) is transmitted by a partially reflective mirror and used to inject the slave (see below). The main beam is sent through an optical isolator and amplified to $450\:$mW by the TA\footnote{TAs with output powers of several Watts are available if more power is required}. The final stage is the second harmonic generation (SHG) using a bulk lithium triborate (LBO) crystal in an optical ring cavity (finesse $\sim700$, FSR $\sim1\:$GHz). The cavity length is locked through a piezo mirror to the laser frequency by employing the Pound-Drever-Hall scheme \cite{Drever1983} with the necessary sidebands being generated by current modulation. This gives frequency-doubled light at $397\:$nm with up to $120\:$mW output power. There is also a fast-feedback path applying the error signal to the ECDL current via a FET modulation input to lock the frequency to the cavity on short time scales. 

The red injection beam is focused through the AOM in double-pass arrangement. We measure first-order diffraction efficiencies of $27\%$ on the first pass and $31\%$ on the second pass giving $8.4\%$ overall. The zeroth order is blocked by a shutter. Alternatively, the AOM can be switched off and the shutter opened to allow injection without a frequency offset. A neutral density (ND) filter is placed in the beam path to give a similar intensity to the first-order beam and to avoid ASE coming from the slave from being re-injected.


The slave diode is protected by two separate $15\:$dB isolators. The injection beam is introduced through the side port on the first of these. Since the beam profiles of master and slave match closely, no additional beam shaping is done. The slave power is monitored on a photodiode (PD) in a pick-off. To characterize the injection for different injection beam powers, slave currents and temperatures, a power meter is placed in the main beam after the pick-off mirror. We observe regions of stable injection spaced by $20\:$mA of slave current. With temperature these shift by $-10\:$mA$/^\circ$C with respect to the slave current. In order to seed the TA, powers $\geq20\:$mW are needed. We typically run the slave diode at $119\:$mA and $20.4^\circ$C with the available injection power of $400\:\mu$W giving us $\sim 24\:$mW of beam power.

Just as for the master, the doubling cavity on the slave laser needs to be locked to the diode's frequency in order to generate the second harmonic. The sidebands generated by current modulation on the master are inherited by the slave via the optical injection. Therefore the master local oscillator signal is also used to generate the slave error signal which is used for feedback on the slave cavity piezo. On short timescales, the slave inherits the frequency stability provided by the master's fast current lock to its own doubling cavity.

\begin{figure}[t]
\centerline{\includegraphics[width=\linewidth]{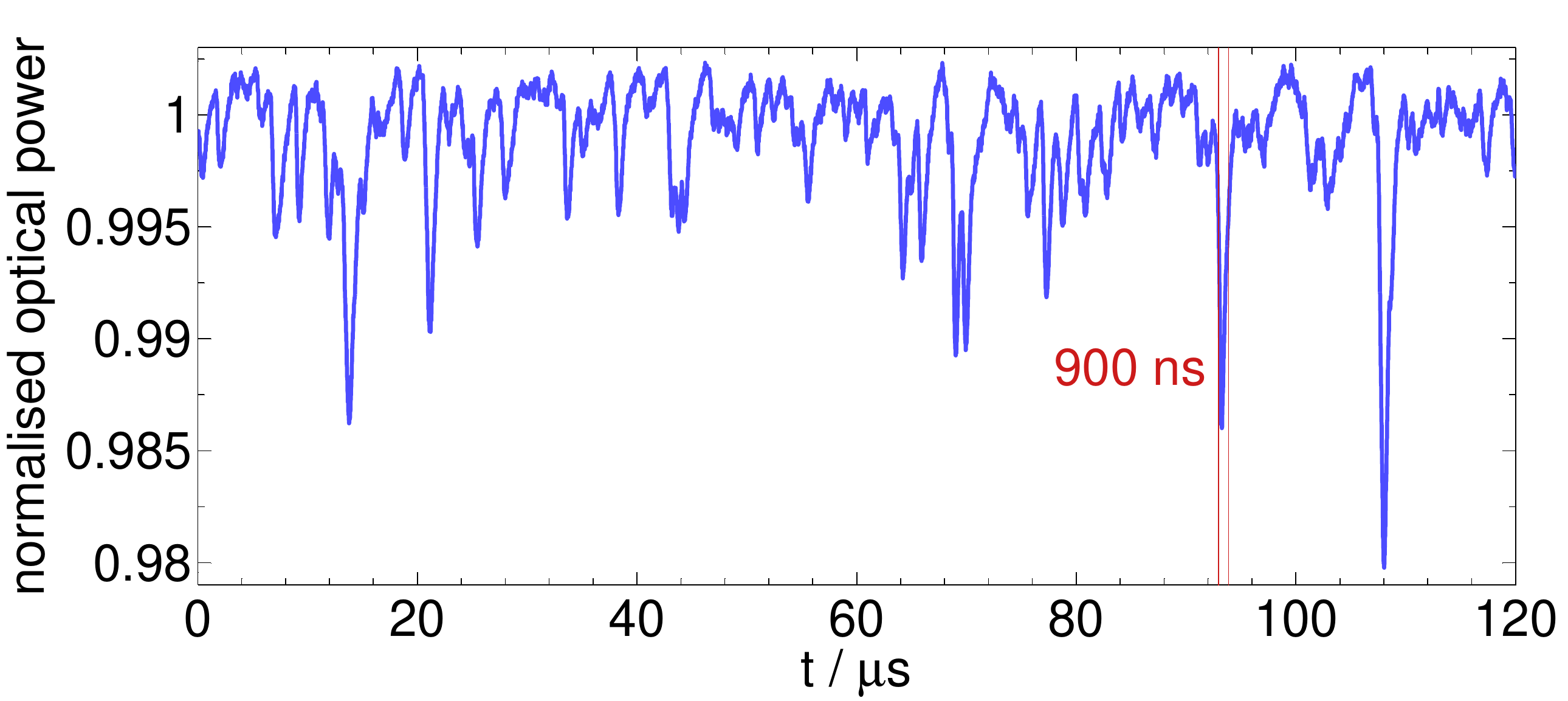}}
\caption{Output power of the SHG cavity vs time. Short ($\sim\,1\:\mu$s) intensity dips on the order of $1$\% can be seen.}
\label{pvst}
\end{figure}

Intensity noise is a disadvantage of using frequency doubling. The intensity of the violet light is monitored on a photo-diode\footnote{Hamamatsu S6775, bandwidth $20\:$MHz}. The signal is processed by a home-built proportional-integral-differential (PID) circuit with a fast (bandwidth $\sim\!5\:$MHz) and a slow (pure integrator) output. The slow signal is used for changing the TA current via the modulation input on the control module (bandwidth $\sim\!10\:$kHz) to cancel drifts in the output power caused for example by cavity misalignment. The fast signal is used to modulate the TA current via a FET with a high bandwidth ($100\:$MHz). The goal is to take out intensity noise caused by vibrations in the cavity or injection path which are too fast to be corrected for by the slow cavity lock. The main limitation to the feedback speed is the cavity transfer function. Any power change before the cavity is only translated into a power change in the violet after a delay on the order of the cavity ring-down time ($0.22\:\mu$s from $90\%-10\%$). Remaining fast intensity noise introduced by the finite quality of the doubling cavity lock and frequency noise on the master diode is shown in figure \ref{pvst}. Typical intensity drop-outs are on the order of $1\:$\% and have a width of around $1\:\mu$s. Assuming a $150\:$kHz Rabi frequency, we estimate that this will contribute $<10^{-4}$ of error to a single $\pi$-pulse. This noise could be reduced further by a fast electro-optic noise cancellation stage.


When switching between the first and zeroth order of the injection AOM, the system takes around $100\:$ms to relock the doubling cavity. This makes it too slow to be used within a single experimental sequence (due to the typical $\sim\!1\:$ms coherence time of a $^{40}$Ca$^+$ qubit), but convenient to switch between experimental runs.

\begin{figure}[t]
\centerline{\includegraphics[width=\linewidth]{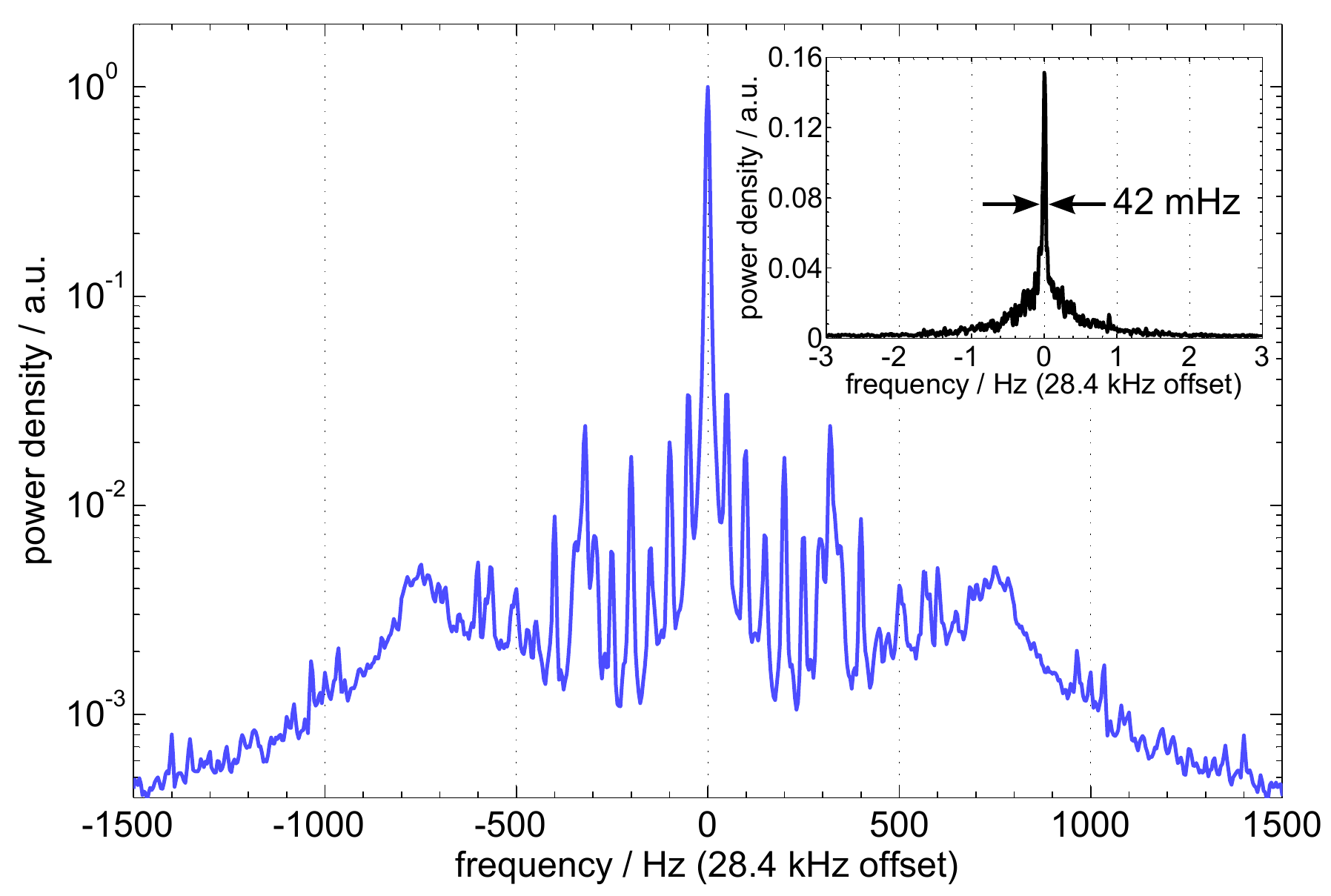}}
\caption{Beat signal from heterodyning the two frequency-doubled beams (mixed down from $3.2\:$GHz) on a log scale ($4\:$Hz per point, acquisition time $250\:$ms, averaged over 50 scans). Servo-bumps from the SHG cavity lock at $\pm750\:$Hz and $50\:$Hz noise peaks are visible. {\bf Inset:} Central peak on a linear scale, FWHM is $42\:$mHz ($7.8\:$mHz per point, acquisition time $128\:$s, averaged over 6 scans).}
\label{beatsignal}
\end{figure}

The relative frequency stability of the frequency-doubled output beams was measured in an optical heterodyne measurement. The violet beams with an offset of $3.2\:$GHz were superimposed on a photodiode\footnote{New Focus 1437} with $25\:$GHz bandwidth placed after $2\:$m of beam path and a $1\:$m optical fiber in each to stabilize beam pointing. The PD signal has a beat note at $3.2\:$GHz which was mixed down to $28.4\:$kHz and analyzed with an FFT signal analyzer\footnote{SRS SR785}. The result is shown in figure \ref{beatsignal}. Apart from the narrow central peak, two servo-bumps from the doubling-cavity lock at $\pm750\:$Hz and a series of peaks at multiples of the $50\:$Hz mains frequency are visible. A close scan around the central peak is shown in the inset. It reveals the width of the beat signal (FWHM) to be $42\:$mHz. The dominant source of noise limiting this width are vibrations in the optical beam path of the interferometer (area $\sim1.3\:$m$^2$) formed by the two beams. The lasers allow us to drive Raman transitions with $1\:$MHz Rabi frequency at a detuning of $\Delta=-440\:$GHz which corresponds to a photon scattering error of $\epsilon=10^{-4}$ for a $\pi$ transition of the clock state qubit \cite{Ozeri07}.

The scheme presented here provides a pair of Raman laser beams for quantum information processing which allows the use of high detuning and hence minimizes photon scattering errors. This is a prerequisite for high-fidelity Raman laser gates in trapped ions.

\end{document}